# Relative Resolution:

# A Computationally Efficient Implementation in LAMMPS

*Mark Chaimovich[1a] and Aviel Chaimovich[2b]*

[1] *Russian School of Mathematics, North Bethesda, Maryland 20852*

[2] *Department of Chemical and Biological Engineering, Drexel University, Philadelphia, Pennsylvania 19104*

[a] mark.chaimovich@russianschool.com
[b] aviel.chaimovich@drexel.edu

## Abstract

Recently, a novel type of a multiscale simulation, called Relative Resolution (RelRes), was introduced. In a single system, molecules switch their resolution in terms of their relative separation, with near neighbors interacting via fine-grained potentials yet far neighbors interacting via coarse-grained potentials; notably, these two potentials are analytically parameterized by a multipole approximation. This multiscale approach is consequently able to correctly retrieve across state space, the structural and thermal, as well as static and dynamic, behavior of various nonpolar mixtures. Our current work focuses on the practical implementation of RelRes in LAMMPS, specifically for the commonly used Lennard-Jones potential. By examining various correlations and properties of several alkane liquids, including complex solutions of alternate cooligomers and block copolymers, we confirm the validity of this automated LAMMPS algorithm. Most importantly, we demonstrate that this RelRes implementation gains almost an order of magnitude in computational efficiency, as compared with conventional simulations. We thus recommend this novel LAMMPS algorithm for anyone studying systems governed by Lennard-Jones interactions.

# 1  Introduction

Ever since the advent of molecular simulations, a main goal of the scientific community has been enhancing their computational efficiency, and this has been frequently done by modifying the interactions between the molecules.[1-4] These days, multiscale algorithms have become especially popular for such a task: At the most basic level, these schemes link together detailed fine-grained (FG) molecular models and simplified coarse-grained (CG) molecular models.[5-6] While there have been various recipes for combining the two models in a single system,[7-16] formulations which switch between the FG and CG interactions in terms of a spatial variable have been particularly common.[17-26]

In this work, we specifically focus on the multiscale framework which we call Relative Resolution (RelRes).[23-26] The main signature of RelRes is that the molecular model is a function of the relative separation: Given two arbitrary molecules, they are modeled via FG interactions if near to each other, yet via CG interactions if far from each other. Such an idea was introduced by Izvekov and Voth in molecular simulations of water,[23] whilst Shen and Hu employed a reminiscent approach in biomolecular systems.[24] Most importantly, while other multiscale approaches typically employ a numerical parametrization between the FG and CG models,[27-28] Chaimovich at al. discovered a crucial benefit of RelRes: It naturally possesses an analytical relation between the FG and CG potentials, stemming in a multipole approximation.[25-26] In fact, by correspondingly employing tabulated potentials in GROMACS,[29] it was shown that RelRes overcomes the transferability and representability issues that hinder most multiscale protocols[30-31]: In particular for several nonpolar mixtures, RelRes correctly captures across state space numerous structural correlations and thermal properties of static and dynamic behavior.[25-26]

In principle, employing RelRes for molecular simulations can yield a significant gain in computational efficiency. Foremost in preparing a multiscale system, the proposed parametrization from the FG model to the CG model can be done analytically on a paper sheet rather than numerically on a computer cluster; this already saves many days of work. Regarding the multiscale simulation itself, we also expect a computational speedup: Although RelRes maintains all degrees of freedom, it can significantly reduce the number of pairwise interactions. Compared with conventional simulations (which have just FG and no CG interactions), RelRes has a similar number of near neighbors but a modest number of far neighbors. While this computational efficiency is expected for RelRes, it has not yet been unequivocally demonstrated.

In this work, we most notably show that if a molecular simulation is implemented via RelRes, it gains almost an order of magnitude in computational efficiency (as compared with a conventional simulation). We in fact show this computational enhancement by implementing RelRes specifically for the Lennard-Jones (LJ) potential in the computational package of LAMMPS.[32] The LJ potential is frequently used by the LAMMPS community in various applications,[33-38] especially in studying elementary alkanes, as well as oligomeric and polymeric hydrocarbons.[39-44] The main benefit of LAMMPS for RelRes is that it inherently supports multiple neighbor lists (based on various cutting distances), and thus, the computational efficiency for RelRes is successfully achieved. Ultimately, we exemplify our LAMMPS algorithm on several alkane liquids (propane, neopentane, alternate cooligomer, block copolymer, etc.). Our current work thus permits the scientific community for the automated use of RelRes in LAMMPS, for the purpose of efficiently studying any nonpolar system.

## 2  Implementation

As mentioned earlier, the main goal of our work is implementing RelRes in LAMMPS,[32] specifically for the LJ potential. Here, we foremost review the theoretical foundation of the RelRes framework,[25-26] introducing several practical modifications along the way. We also thoroughly describe all the technical aspects of our implementation in LAMMPS.

### 2.1  Theoretical Basis

Consider a molecular system. Naturally, each molecule is composed of several atoms; in this article, we will commonly use the term atom for the concept of a "united" atom (e.g., a principal carbon with its adjacent hydrogens). In the many molecular simulations, each atom is represented by a FG site. It may be convenient to break up each molecule into several groups of atoms. In such cases, each group is represented by a CG site. The distinct aspect of the various multiscale strategies is that the FG and CG models are combined in a single molecular simulation.[17-22] Fig. 1 elaborates on this multiscale concept, showing two arbitrary groups of atoms (not necessarily of the same molecule). The CG sites are labeled with Latin letters $i$ and $j$, and the FG sites are labeled with Greek letters $\mu$ and $\nu$. For brevity, we will often address an arbitrary CG site with a Latin letter $l$, and an arbitrary FG site with a Greek letter $\upsilon$. Note that a main aspect of a group is its "mapping ratio" $n$ (i.e., the number of FG sites mapped onto a CG site). Of course, the mapping ratio can be different for different groups; so in principle, $n$ comes with an index $l$, yet we omit it for clarity.

Intrinsically, FG sites $\mu$ and $\nu$ interact via a potential function $u_{\mu\nu}^{FG}(r_{\mu\nu})$, where $r_{\mu\nu}$ denotes the respective distance between them. Alternatively, CG sites $i$ and $j$ interact with a potential function $u_{ij}^{CG}(r_{ij})$, where $r_{ij}$ denotes the respective distance between them. The unique signature of the RelRes approach, as compared with other multiscale strategies, is that the FG and CG interactions are turned on or off depending on these distances:

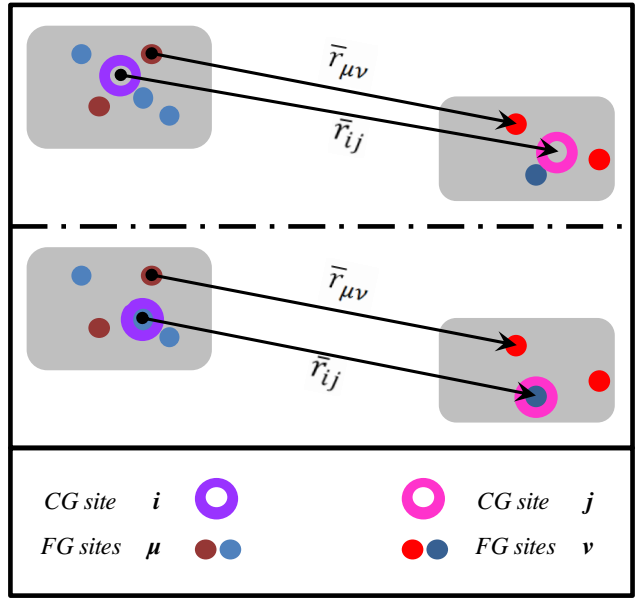

FIG. 1. The topology of two groups of atoms. Their CG sites are represented by hollow rings, while their FG sites are represented by replete disks. The left side is for CG site $i$ with its FG sites $\mu$, and the right side is for CG site $j$ with its FG sites $\nu$. In the upper portion, the CG site is located at the center of mass between the FG sites and the lower portion, the CG site is instead placed at the same location as one of the FG sites. The various colors of these sites signify that they can all have different interaction parameters. The gray shading does not have much of a physical meaning: it just helps us delineate an effective boundary of a group. The various arrows are distance vectors.

Specifically, nearby sites interact via FG potentials, while faraway sites interact via CG potentials.[23-26] We will now define a distance parameter at which interactions switch between the FG potential and the CG potential, and we will specifically call it the "switching distance" $r_s$; realize that in principle, $r_s$ may be distinct for each molecular pair $ij$, yet we omit these indices for clarity. In the most recent formulation of RelRes,[25-26] its FG potential is defined as

$$\tilde{u}_{\mu\nu}^{FG}(r) = \begin{cases} u_{\mu\nu}^{FG}(r) - u_{\mu\nu}^{FG}(r_s) & if\ r < r_s \\ 0 & if\ r \geq r_s \end{cases}, \quad (1)$$

and its CG potential is defined as

$$\tilde{u}_{ij}^{CG}(r) = \begin{cases} u_{ij}^{CG}(r_s) & if\ r < r_s \\ u_{ij}^{CG}(r) & if\ r \geq r_s \end{cases}. \quad (2)$$

In essence, these are just Heaviside functions, with the constants $u_{\mu\nu}^{FG}(r_s)$ and $u_{ij}^{CG}(r_s)$ respectively providing continuity of the FG and CG potentials at $r_s$. The total energy of the system can be expressed by combining Eqs. (1) and (2), while performing the appropriate summation over all sites

$$\widetilde{U} = \frac{1}{2} \sum_{i \neq j} \left[ \sum_{\mu\nu} \tilde{u}_{\mu\nu}^{FG}(r_{\mu\nu}) + \tilde{u}_{ij}^{CG}(r_{ij}) \right]. \tag{3}$$

Notice that with $r_s \to \infty$, we obtain a pure FG system, and with $r_s \to 0$, we obtain a pure CG system. We will call the former the "reference" system, as this is frequently the benchmark used in molecular simulations (i.e., it supposedly describes the correct behavior of a liquid of interest).

In order to ensure a correct representation of the system behavior, multiscale strategies require appropriate parametrization between the FG and CG models. To perform such a parametrization, different numerical route can be employed.[27-28,45-46] In such established procedures, the parametrization is notably performed across the entire domain of the pairwise distance. While the initial efforts in RelRes essentially used such strategies,[23-24] the most recent RelRes framework found a natural way of connecting the FG and CG potentials.[25-26] Above all, RelRes only cares about replacing the FG potential with the CG potential beyond $r_s$. Thus, RelRes proceeds by equating the FG and CG energies at the "infinite limit" of the relative separation between an isolated molecular pair (i.e., $r \gg |\bar{r}_{ij} - \bar{r}_{\mu\nu}|$); this yields the following analytical parametrization[25]

$$u_{ij}^{CG}(r) = \sum_{\mu\nu} u_{\mu\nu}^{FG}(r). \tag{4}$$

With this parametrization, it is obvious that the RelRes potential defined by Eq. (3) will approximately capture the behavior of the reference system if for all sites $r_s \gg |\bar{r}_{ij} - \bar{r}_{\mu\nu}|$ (i.e., the dimension of each group of atoms is relatively small, in comparison with the switching distance); this is in fact why we often partition a single molecule into several groups. Note that the formal derivation of this parametrization proceeds via a Taylor series.[26] Here, we present just the zero-order term of this multipole approximation, which is sufficient for nonpolar molecules; for polar scenarios, the zero-order term may vanish, and thus, the first-order or second-order terms become essential.

In Refs. 25-26, the CG site was placed at the center of mass given by the FG sites, which required the construction of a virtual site; such mapping is depicted in the top portion of Fig.1. However, we are not interested in exact trajectories in molecular simulations, and thus, the placement at the center of mass is not an absolute requirement. In this current work, we place the CG site on one of its FG sites, which logically should be the site closest to the center of mass; such mapping is depicted in the bottom portion of Fig. 1. Above all, the need for an extra (virtual) site is eliminated.

We call the site, in which the CG and FG potentials merge, "hybrid". Conversely, we call all other sites "ordinary". Notably, hybrid sites play a dual function, interacting via a FG potential with near neighbors (even ordinary ones) and via a CG potential with far neighbors (just hybrid ones). Conversely, ordinary sites interact only via a FG potential just with near neighbors (there are no sites that interact only via a CG potential). Observe here that an arbitrary group of atoms will be comprised of one hybrid site and $n-1$ ordinary sites. Now for completeness, we can define the CG interactions for all site combinations in the following way:

$$u_{\mu\nu}^{CG}(r) = \begin{cases} u_{ij}^{CG}(r) & \mu \wedge \nu \text{ are hybrid} \\ 0 & \mu \vee \nu \text{ is ordinary} \end{cases} \quad (5)$$

The definition for the FG interactions, for both hybrid and ordinary sites, is the same. Thus, we can combine Eqs. (1) and (2) in the following manner

$$\tilde{u}_{\mu\nu}(r) = \begin{cases} u_{\mu\nu}^{FG}(r) - u_{\mu\nu}^{FG}(r_s) + u_{\mu\nu}^{CG}(r_s) & if\ r < r_s \\ u_{\mu\nu}^{CG}(r) & if\ r \geq r_s \end{cases}. \quad (6)$$

For clarity, throughout this article, we often omit the indices, having $u(r) = u_{\mu\nu}(r)$ or $\tilde{u}(r) = \tilde{u}_{\mu\nu}(r)$. However, keep in mind that different sites can have their own unique potential.

## 2.2 Practical Modifications

Naturally, molecular simulations deal with interactions of a finite range, and thus, we must introduce a cutting distance $r_c$ in the RelRes potential of Eq. (6) (i.e., $\tilde{u}(r) = 0$ for $r > r_c$). Note that similar with $r_s$, $r_c$ can be in principle different for different groups, and of course, $r_c > r_s$.

Importantly, most packages, including LAMMPS, are based on Newtonian motion, and therefore, from a practical perspective, Eq. (6) must be modified, so that there is no singularity in the force (i.e., $-\tilde{u}'(r)$) at $r_s$, as well as at $r_c$. For this purpose, we define $r_{s\pm} = r_s \pm \frac{1}{2}\Delta r_s$: the FG potential is only present for $r < r_{s-}$, and the CG potential is only present for $r \geq r_{s+}$. In the interval $[r_{s-}, r_{s+})$ (i.e, the smoothing zone), the potential is governed by the smoothing function. In an analogous manner, a smoothing function is also applied at the cutting distance, with its inner and outer values defined by $r_{c\pm} = r_c \pm \frac{1}{2}\Delta r_c$. Note that we often omit the indices $s$ and $c$ (e.g., $\Delta r = r_+ - r_-$), if our discussion is relevant for both smoothing zones.

In principle, the smoothing zone can have any functionality; it is just constrained by the appropriate application of the boundary conditions at, $r_\pm$, which at the most basic level, just ensure the continuity of the energy, $\tilde{u}(r)$, together with the force, $-\tilde{u}'(r)$. For enhanced smoothness, we also apply an additional continuity requirement for the ensuing derivative, $-\tilde{u}''(r)$. We will specifically cast each smoothing function in terms of the relative location within each smoothing zone, $\delta = r - r_-$, which in turn means that $0 \leq \delta \leq \Delta r$.

In practice, we have three smoothing functions. Reminiscent of the familiar situation, we have two smoothing functions which shut down interactions: $f_{s-}^{FG}(\delta)$ turns off the FG force at $r_{s-}$, smoothing it to zero at $r_{s+}$, and $f_{c-}^{CG}(\delta)$ turns off the CG force at $r_{c-}$, smoothing it to zero at $r_{c+}$. Notably, we also have one smoothing function which starts up an interaction: $f_{s+}^{CG}(\delta)$ turns on the CG force at $r_{s+}$, smoothing it from zero at $r_{s-}$. Note that in our notation, a fourth smoothing function in principle exists: $f_{c-}^{FG}(\delta)$ does nothing for the FG force at $r_{c-}$ (it is formally zero across its entire domain).

For all these functions, we have five boundary conditions. Four of the conditions take care of the first and second derivatives of the potential at both boundaries, $r_-$ and $r_+$. The fifth condition takes care of the continuity of the potential at one of the boundaries. Note that no sixth condition is required since the shifting constants of Eq. (6) take care of the continuity of the potential at the other boundary.

The boundary conditions for $f_{s-}^{FG}$ and $f_{c-}^{CG}$ are very similar, while the boundary conditions for $f_{s+}^{CG}$ are slightly different. In the following definitions of the boundary conditions, we omit their FG and CG superscripts for clarity (as is often done throughout the manuscript, we also omit the $s$ and $c$ of the subscripts). For $f_{s-}^{FG}$ and $f_{c-}^{CG}$, we have:

$$\begin{aligned}
f_-(0) &= u(r_-) \\
f'_-(0) &= u'(r_-) \\
f''_-(0) &= u''(r_-) \\
f'_-(\Delta r) &= 0 \\
f''_-(\Delta r) &= 0
\end{aligned} \qquad (7)$$

and for $f_{s+}^{CG}$, we have:

$$\begin{aligned}
f_+(\Delta r) &= u(r_+) \\
f'_+(\Delta r) &= u'(r_+) \\
f''_+(\Delta r) &= u''(r_+) \\
f'_+(0) &= 0 \\
f''_+(0) &= 0
\end{aligned} \qquad (8)$$

In turn, the shifting constants guaranteeing the continuity of the potential on the other boundary of the smoothing zone will be

$$\begin{aligned}
\Gamma_{c-} &= f_{c-}^{CG}(\Delta r_c), \\
\Gamma_{s+} &= f_{c-}^{CG}(\Delta r_c) + f_{s-}^{FG}(\Delta r_s), \\
\Gamma_{s-} &= f_{c-}^{CG}(\Delta r_c) + f_{s-}^{FG}(\Delta r_s) - f_{s+}^{CG}(0).
\end{aligned} \qquad (9)$$

Note that while $\Gamma_{c+}$ in principle exists, it is formally identical with $\Gamma_{c-}$.

Now we are ready to formulate our entire RelRes potential for practical use in molecular simulations:

$$\tilde{u}(r) = \begin{cases} u^{FG}(r) - \Gamma_{s-} & \text{if } r \in [0, r_{s-}), \\ f_{s-}^{FG}(r - r_{s-}) + f_{s+}^{CG}(r - r_{s-}) - \Gamma_{s+} & \text{if } r \in [r_{s-}, r_{s+}), \\ u^{CG}(r) - \Gamma_{c-} & \text{if } r \in [r_{s+}, r_{c-}), \\ f_{c-}^{CG}(r - r_{c-}) - \Gamma_{c-} & \text{if } r \in [r_{c-}, r_{c+}), \\ 0 & \text{if } r \in [r_{c+}, \infty). \end{cases} \qquad (10)$$

## 2.3 Lennard-Jones Potential

The basis for most molecular simulations is the LJ potential. In general, it is perfectly sufficient for nonpolar molecules, and in conjunction with the Coulomb potential, it can adequately describe polar molecules as well. Indeed, the LAMMPS package supports multiple variations of the LJ potential. Therefore, we consider here the utilization of the LJ potential with RelRes.

For the interaction between FG sites $\mu$ and $\nu$, the LJ potential can be expressed as

$$u_{\mu\nu}^{FG}(r) = 4\epsilon_{\mu\nu}^{FG}\left[\left(\frac{\sigma_{\mu\nu}^{FG}}{r}\right)^{12} - \left(\frac{\sigma_{\mu\nu}^{FG}}{r}\right)^{6}\right] \tag{11}$$

with $\sigma_{\mu\nu}^{FG}$ and $\epsilon_{\mu\nu}^{FG}$ being its respective length and energy parameters. Substituting this expression in Eq. (4), we obtain that the interaction between CG sites $i$ and $j$ is governed by a similar expression

$$u_{ij}^{CG}(r) = 4\epsilon_{ij}^{CG}\left[\left(\frac{\sigma_{ij}^{CG}}{r}\right)^{12} - \left(\frac{\sigma_{ij}^{CG}}{r}\right)^{6}\right] \tag{12}$$

where

$$\sigma_{ij}^{CG} = \frac{\left(\sum_{\mu\nu}\epsilon_{\mu\nu}^{FG}\sigma_{\mu\nu}^{FG\,12}\right)^{1/6}}{\left(\sum_{\mu\nu}\epsilon_{\mu\nu}^{FG}\sigma_{\mu\nu}^{FG\,6}\right)^{1/6}}, \qquad \epsilon_{ij}^{CG} = \frac{\left(\sum_{\mu\nu}\epsilon_{\mu\nu}^{FG}\sigma_{\mu\nu}^{FG\,6}\right)^{2}}{\left(\sum_{\mu\nu}\epsilon_{\mu\nu}^{FG}\sigma_{\mu\nu}^{FG\,12}\right)}. \tag{13}$$

Eq. (13) gives us a simple analytical relation between FG and CG parameters.

Now we need to define the CG interactions for all site combinations in a similar manner as done in Eq. (5). Specifically for the LJ potential it is fulfilled by setting

$$\sigma_{\mu\nu}^{CG} = \sigma_{ij}^{CG}, \qquad \epsilon_{\mu\nu}^{CG} = \begin{cases} \epsilon_{ij}^{CG} & \mu \wedge \nu \text{ are hybrid} \\ 0 & \mu \vee \nu \text{ is ordinary} \end{cases} \tag{14}$$

Having these relations in mind, while respectively replacing $u^{FG}$ and $u^{CG}$ in Eq. (10) with Eqs. (11) and (12) we attain the expression for RelRes with the LJ potential.

The parameters between the same FG sites, $\sigma$ and $\epsilon$, are often known. In turn, the mixed parameters between arbitrary sites can be determined via a geometric mean: $\epsilon_{\mu\nu} = \sqrt{\epsilon_\mu \epsilon_\nu}$ and $\sigma_{\mu\nu} = \sqrt{\sigma_\mu \sigma_\nu}$. We can then simplify Eq. (13) for CG site $l$ comprised from FG sites $\upsilon$ to obtain

$$\sigma_l^{CG} = \frac{\left(\sum_\upsilon \sqrt{\epsilon_\upsilon^{FG}\sigma_\upsilon^{FG\,12}}\right)^{1/3}}{\left(\sum_\upsilon \sqrt{\epsilon_\upsilon^{FG}\sigma_\upsilon^{FG\,6}}\right)^{1/3}}, \qquad \epsilon_l^{CG} = \frac{\left(\sum_\upsilon \sqrt{\epsilon_\upsilon^{FG}\sigma_\upsilon^{FG\,6}}\right)^{4}}{\left(\sum_\upsilon \sqrt{\epsilon_\upsilon^{FG}\sigma_\upsilon^{FG\,12}}\right)^{2}} \tag{15}$$

and subsequently, the geometric mean holds here as well, $\epsilon_{ij}^{CG} = \sqrt{\epsilon_i^{CG}\epsilon_j^{CG}}$ and $\sigma_{ij}^{CG} = \sqrt{\sigma_i^{CG}\sigma_j^{CG}}$. To summarize, for the LJ potential Eqs. (13) and (15) are very convenient: They determine the CG parameters analytically, with no need for an actual molecular simulation; only knowledge of the FG parameters of the reference system is required. Eq. (13) can be always employed, but if geometric mixing is applicable in the LJ potential, Eq. (15) is the relevant one for use in molecular simulations.

## 2.4 Polynomial Smoothing

Let us now discuss our smoothing functions. Recall that RelRes employs three such functions (i.e., $f_{s-}^{FG}$, $f_{s+}^{CG}$, and $f_{c-}^{CG}$). In our arguments, we may omit superscripts for clarity. In the spirit of LAMMPS, we choose a polynomial for the smoothing function

$$f_{\pm}(\delta) = \sum_{m=0}^{4} \gamma_{m\pm}(\delta)^m, \tag{16}$$

where the five coefficients $\gamma_{m\pm}$ can ensure the continuity of the potential with its first and second derivatives, and again, $\delta$ is the relative location within the smoothing zone. Applying the boundary conditions of Eqs. (7) and (8), we obtain a system of liner equations for each smoothing function, and in turn we get the following

$$\begin{aligned}
\gamma_{0-} &= u(r_-) \\
\gamma_{1-} &= u'(r_-) \\
\gamma_{2-} &= \frac{1}{2}u''(r_-) \\
\gamma_{3-} &= -\frac{u'(r_-)}{\Delta r^2} - \frac{2u''(r_-)}{3\Delta r} \\
\gamma_{4-} &= \frac{u'(r_-)}{2\Delta r^3} + \frac{u''(r_-)}{4\Delta r^2}
\end{aligned} \tag{17}$$

and

$$\begin{aligned}
\gamma_{0+} &= u(r_+) - \frac{1}{2}u'(r_+)\Delta r + \frac{1}{12}u''(r_+)\Delta r^2 \\
\gamma_{1+} &= 0 \\
\gamma_{2+} &= 0 \\
\gamma_{3+} &= \frac{u'(r_+)}{\Delta r^2} - \frac{u''(r_+)}{3\Delta r} \\
\gamma_{4+} &= -\frac{u'(r_+)}{2\Delta r^3} - \frac{u''(r_+)}{4\Delta r^2}.
\end{aligned} \tag{18}$$

Besides, with this notation, the shifting constants of Eq. (9) can be expressed as

$$\Gamma_{c-} = \sum_{m=0}^{4} \gamma_{m-}^{CG}(\Delta r_c)^m$$

$$\Gamma_{s+} = \sum_{m=0}^{4} \gamma_{m-}^{CG}(\Delta r_c)^m + \sum_{m=0}^{4} \gamma_{m-}^{FG}(\Delta r_s)^m \qquad (19)$$

$$\Gamma_{s-} = \sum_{m=0}^{4} \gamma_{m-}^{CG}(\Delta r_c)^m + \sum_{m=0}^{4} \gamma_{m-}^{FG}(\Delta r_s)^m - \gamma_{0+}^{CG}$$

## 2.5 LAMMPS Code

The introduction of RelRes into the LAMMPS package requires computing pairwise interactions according to the RelRes potential of Eq. (10). To specifically use this equation with the LJ potential, we need to substitute Eqs. (11) and (12) in place of the general $u(r)$ of Eq. (10). Besides, we substitute the polynomial smoothing from Eq. (16) in place of the general $f(\delta)$. In turn, we obtain the following piecewise potential:

$$\tilde{u}(r) = \begin{cases} 4\epsilon^{FG}\left[\left(\frac{\sigma^{FG}}{r}\right)^{12} - \left(\frac{\sigma^{FG}}{r}\right)^{6}\right] - \Gamma_{s-} & \text{if } r \in [0, r_{s-}), \\ \sum_{m=0}^{4}(\gamma_{m-}^{FG} + \gamma_{m+}^{CG})(r - r_{s-})^m - \Gamma_{s+} & \text{if } r \in [r_{s-}, r_{s+}), \\ 4\epsilon^{CG}\left[\left(\frac{\sigma^{CG}}{r}\right)^{12} - \left(\frac{\sigma^{CG}}{r}\right)^{6}\right] - \Gamma_{c-} & \text{if } r \in [r_{s+}, r_{c-}), \\ \sum_{m=0}^{4}\gamma_{m-}^{CG}(r - r_{c-})^m - \Gamma_{c-} & \text{if } r \in [r_{c-}, r_{c+}), \\ 0 & \text{if } r \in [r_{c+}, \infty), \end{cases} \qquad (20)$$

where the coefficients of the smoothing functions $\gamma_{m\pm}$ are defined by Eqs. (17) and (18), whilst the shifting constants $\Gamma_{\pm}$ are defined by Eq. (19). Note here that in view of Eq. (14), the last two cases of Eq. (20) (i.e., $r \in [r_{s+}, r_{c+})$) are only applicable for hybrid sites, while the initial two cases (i.e., $r \in [0, r_{s+})$) are completely applicable for all sites. Fig. 2 shows two variations of this RelRes interaction between hybrid sites; each one has its specific set of parameters, most notably differing in their mapping ratio, as well as in their switching distance: In particular, $n = 3$ with $r_s = 0.6$nm is in blue, and $n = 5$ with $r_s = 0.7$nm. is in red. The potential is plotted in the top panel, while the force is plotted in the bottom panel.

The LAMMPS package itself is designed in a modular approach such that it allows for the easy addition of a new functionality via a new C++ class, while interfacing with the rest of the code (e.g., time integration options, error diagnostic, neighbor list management, input-output functions, etc.). Therefore, behind each pairwise interaction function in LAMMPS, there is a separate C++ class that governs the calculation of potentials along with their associated forces. There are already multiple existing built-in classes in LAMMPS that support various LJ pair interactions. We take the `pair_lj_smooth` class with its `lj/smooth` pair style, as a template for our RelRes implementation: We name our new class `pair_lj_relres` with its new pair style `lj/relres`.

Each pair style in LAMMPS uses a specific set of parameters, controlled by special commands in an input script. The first command is `pair_style`, whose leading argument is the name of the particular interaction which is used (e.g., `lj/smooth`, `lj/relres`, etc.). The complementary arguments which come with `pair_style` are typically global, being applied on the entire system (in the case of `lj/smooth`, these arguments are the global inner and outer cutting distances). The second command is `pair_coeff`, whose leading two arguments are the identifiers of a pair of atom types (in LAMMPS, an atom type defines a set of atoms with identical characteristics). The complementary arguments of `pair_coeff` are the interaction parameters of the specified pair of atom types (in the case of `lj/smooth`, these arguments are the energy and length scales of the LJ potential). Note here that the interaction parameters must be defined for all atom types, and thus, it generally requires multiple `pair_coeff` commands. Besides, the `pair_coeff` command provides an option to override the global values defined by the `pair_style` command (in the case of `lj/smooth`, these arguments are the specific inner and outer cutting distances for the specified pair of atom types).

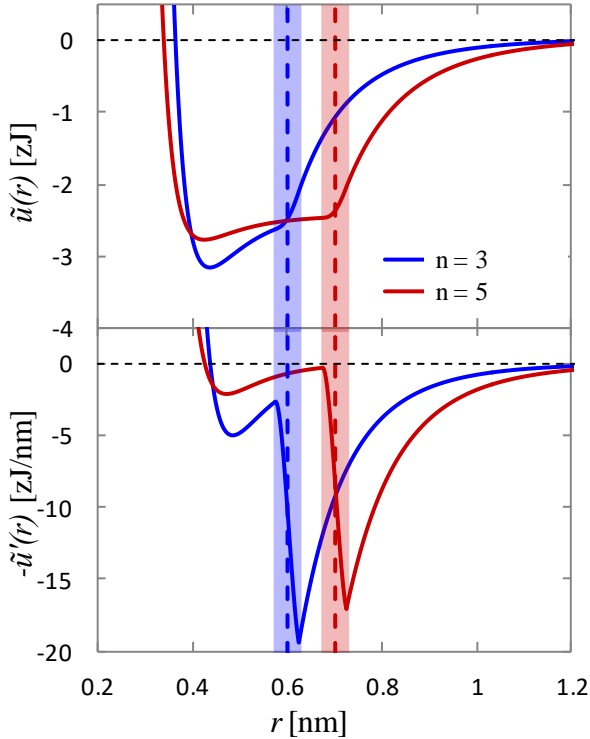

FIG. 2. The RelRes version of the LJ interaction between a pair of hybrid sites. While the defining potential is shown in the top panel, the corresponding force is shown in the bottom panel. There are actually two separate examples here: The blue curves represent our ideal model for propane (i.e., a mapping ratio of $n = 3$, with a switching distance of $r_s = 0.6$nm), and the red curves represent our ideal model for neopentane (i.e., a mapping ratio of $n = 5$, with a switching distance of $r_s = 0.7$nm). Each dashed vertical line corresponds with the location of the switching distance, while its shaded region represents the smoothing zone. The mapping for these alkanes is described in Fig. 3, and their LJ parameters are specifically given in Table 1.

Analogous with Eq. (20), `lj/relres` requires eight parameters: Besides the four LJ energy and length parameters of the FG and CG potentials (i.e., $\epsilon^{FG}, \sigma^{FG}, \epsilon^{CG}, \sigma^{CG}$), there are the four distance parameters of the smoothing zones (i.e., $r_{s-}, r_{s+}, r_{c-}, r_{c+}$). In our implementation, the distance parameters are defined globally in the `pair_style` command; there is an option of overriding them for a particular atom type with the `pair_coeff` command. On the other hand, since the LJ energy and length parameters are specific for each atom type, they are defined in the `pair_coeff` command. Here is the formal syntax of these commands in the input script:

```
pair_style  lj/relres  r_s-  r_s+  r_c-  r_c+
pair_coeff  α  β  ε^FG_αβ  σ^FG_αβ  ε^CG_αβ  σ^CG_αβ  [ r_s-  r_s+  r_c-  r_c+ ]
```

where $\alpha$ and $\beta$ are atom types, and brackets denote optional arguments. Keep in mind that these commands require $r_{s-} \leq r_{s+} \leq r_{c-} \leq r_{c+}$; if desired, one may eliminate either smoothing zone by accordingly setting $r_- = r_+$.

Of course, the flexibility of LAMMPS allows placing any values for the parameters in the input script; we now delineate the correct use of RelRes in LAMMPS. Foremost, one must presume a set of parameters for the FG model that applies for all atom types (for example, by invoking OPLS_UA[47]). If this FG set complies with geometric mixing, the CG parameters of the hybrid sites should be calculated using Eq. (15); otherwise, Eq. (13) should be used instead, and the `pair_coeff` command should be explicitly defined for all combinations of atom types $\alpha \neq \beta$. On the other hand for the ordinary sites, the CG energy parameter (not necessarily the CG length parameter) must be set to zero according to Eq. (14). The distance parameters are also crucial for the correct performance of RelRes. All atom types belonging to a specific group must have the same $r_\pm$ arguments. While the cutting distance $r_{c\pm}$ must be the same as that of the reference system, the ideal choice for the switching distance $r_{s\pm}$ is thoroughly discussed later in the publication.

For various modifications of the interactions, the `pair_modify` command is often used in LAMMPS. RelRes particularly supports two keywords of this command. Firstly, the `shift` keyword corresponds with the shifting constant $\Gamma_{c-}$ (the other shifting constants $\Gamma_{s\pm}$ are not impacted by the `shift` keyword): A value of `yes` guarantees that Eq. (19) is satisfied, while the default value of `no` sets $\Gamma_{c-}$ to zero. Secondly, the `mix` keyword is responsible for the mixing rules between atom types $\alpha \neq \beta$ (if `pair_coeff` is not explicitly defined for them): The default value is `geometric`, yet if desired, the other typical mixing rules of LAMMPS are also available.

As an illustration, let us define a hypothetical system which has a group with two atom types: The first type corresponds to hybrid sites, and the second type corresponds to ordinary sites. Here is the input script relevant for such a supposed system:

```
pair_style lj/relres 0.68e-9  0.72e-9  1.20e-9  1.40e-9    #Line1
pair_coeff 1   1      0.80e-21 0.37e-9 14.9e-21 0.39e-9    #Line2
pair_coeff 2   2      1.00e-21 0.40e-9 0.00     0.39e-9    #Line3
pair_modify shift yes                                       #Line4
```

Everything is presented here in SI units, with $10^{-9}$m = 1nm and $10^{-21}$J = 1zJ. For convenience, we numbered the command lines; as usual, a text following a pound sign is processed as a comment. `Line1` sets the global $r_{s\pm} = 0.70 \pm 0.02$nm and the global $r_{c\pm} = 1.30 \pm 0.10$nm. `Line2` defines the interaction parameters between the hybrid sites with $\epsilon_{11}^{FG} = 0.35$zJ, $\sigma_{11}^{FG} = 0.38$nm, $\epsilon_{11}^{CG} = 21.1$zJ, and $\sigma_{11}^{CG} = 0.39$nm. `Line3` defines the interaction parameters between the ordinary sites with $\epsilon_{22}^{FG} = 1.01$zJ, $\sigma_{22}^{FG} = 0.40$nm, $\epsilon_{22}^{CG} = 0.00$zJ, and $\sigma_{22}^{CG} = 0.39$nm. `Line4` provides the appropriate calculation for the shifting constant $\Gamma_{c-}$ (use of the `mix` keyword is unnecessary as `geometric` is the default value in LAMMPS). Note that `Line2` and `Line3` in the example above only have four (mandatory) arguments, implying that the global $r_\pm$ of `Line1` are used for these atom types. If some of the global $r_\pm$ parameters should be overridden for this group, four other (optional) parameters must be included. Overriding the switching distance with $r_{s\pm} = 0.60 \pm 0.02$nm is illustrated in the example below:

```
pair_coeff 1  1      0.80e-21 0.37e-9 14.9e-21 0.39e-9   &
                     0.58e-9  0.62e-9 1.20e-9  1.40e-9   #Line2
pair_coeff 2  2      1.00e-21 0.40e-9 0.00     0.39e-9   &
                     0.58e-9  0.62e-9 1.20e-9  1.40e-9   #Line3
```

Note that the ampersand sign indicates the continuation of a command onto the next line.

In addition, the flexibility of LAMMPS allows employing the `lj/relres` style as one of the subtypes within the `hybrid/overlay` (also `hybrid`) pair style, in turn, enabling its use with multiple other potentials in a single molecular simulation. Most notably, it can be combined with the Coulomb potential in studying polar systems. Here is an illustration of a possible combination:

```
pair_style hybrid/overlay lj/relres rs- rs+ rc- rc+ coul/cut rc
pair_coeff α β lj/relres ε_{αβ}^{FG} σ_{αβ}^{FG} ε_{αβ}^{CG} σ_{αβ}^{CG} [ rs- rs+ rc- rc+ ]
pair_coeff α β coul/cut
```

Here, a name of a pair style appears as an argument in the `pair_coeff` command, allowing for the definition of multiple potentials for the same pair of atom types. Note that these other potentials maintain their FG nature, and thus, no CG treatment is required for these interactions.

## 2.6 Computational Efficiency

Calculation of pairwise interactions is the main consumption factor in molecular simulations: Usually, these contribute to more than 90% of the CPU time. Due to a reduced number of pairwise interactions, CG models have a computational advantage over FG models, and, therefore, we should expect a similar effect in a multiscale system like RelRes.

LAMMPS, like most other computational packages, uses a neighbor list for calculating pairwise energetics. In general, only sites within a specific distance (slightly beyond the cutting distance) are included in the neighbor list. Importantly, the interactions are computed merely with these relevant neighbors, and therefore, the CPU time is proportional to the characteristic size of the neighbor list, $\mathbb{N}$.

In the RelRes algorithm, the list size depends on the category of the site (an ordinary site or a hybrid site). In either case, the size of the list is significantly reduced as compared with a reference system. Foremost, an ordinary site has less neighbors as its list size is governed by the fairly short $r_s$ rather than the fairly long $r_c$. On the other hand, in consideration of the mapping ratio $n$, a hybrid site has less neighbors because beyond the switching distance, it interacts only with other hybrid sites yet not with any ordinary sites. Therefore, the decrease of $r_s$ and the increase of $n$ are the main factors that can enhance the computational efficiency of RelRes.

By employing the relevant CPU times, let us now define a measure for the computational efficiency of RelRes (as compared with the corresponding reference system). Let $\tau_{\text{tot}}$ be the total CPU time of a molecular simulation, and let $\tilde{\tau}$ be the CPU time taken just by the pairwise interactions; $\tau_{\text{tot}}^*$ and $\tilde{\tau}^*$ are respectively the RelRes times normalized by the corresponding values of the reference system. In turn, the measures of the computational efficiency are respectively the inverses of these normalized times, $\eta_{\text{tot}}^* = 1/\tau_{\text{tot}}^*$ and $\tilde{\eta}^* = 1/\tilde{\tau}^*$: These are natural indicators of how much faster RelRes performs compared with the reference system.

Computation of the pairwise interactions is the only aspect of a molecular simulation changed by RelRes, and since this CPU time is proportional to the list size, we estimate

$$\tilde{\eta}^* \approx 1/\mathbb{N}^*, \qquad (21)$$

where $\mathbb{N}^*$ represents the characteristic size of the RelRes neighbor list, normalized by the corresponding value of the reference system. Since pairwise interactions dominate the CPU consumption, we may subsequently presume

$$\eta_{\text{tot}}^* \approx \chi/\mathbb{N}^*, \qquad (22)$$

where $\chi$ is an empirical parameter that depends on the ratio $\tilde{\tau}/\tau_{\text{tot}}$; when this ratio approaches unity, $\chi$ approaches unity as well, yet $\chi$ is always less than one.

# 3  Verification

We now shift our attention to testing the RelRes algorithm for the LJ potential in LAMMPS. Can it correctly describe the structural and thermal behavior of various liquids, and can it do so with an enhanced computational efficiency? The ultimate goal is to show that our LAMMPS implementation can be effectively used for complex mixtures (solutions with cooligomers, copolymers, etc.), and for achieving this aim, we systematically evolve here simple systems of "monomeric" and "dimeric" liquids. We succeed in this piecemeal task by employing the appropriate switching distance (between the FG and CG potentials), together with the appropriate parametrization of Eq. (15) (for the $\sigma$ and $\epsilon$ parameters). We specifically examine the following alkanes:

- Two distinct monomers, $C_3H_8$ (propane) and $C_5H_{12}$ (neopentane), which form the basis for the remainder of our investigation.
- Two basic dimers, $C_3H_7$-$C_3H_7$ (hexane) and $C_5H_{11}$-$C_5H_{11}$ (bineopentyl), both of which are symmetric extensions of our monomers.
- A dimer $C_3H_7$-$C_5H_{11}$ (2,2-dimethylhexane) which is an asymmetric construction of our monomers.
- A dimer $C_4H_9$-$C_4H_9$ (biisobutyl) which is a symmetric construction from groups, with a mapping ratio of 4, that have not been investigated as monomers.
- An isobutane solution which contains four cooligomeric chains $C_3H_7$-$(C_5H_{10}$-$C_3H_6)_4$-$C_5H_{11}$.
- An isobutane solution which contains two copolymeric chains $C_3H_7$-$(C_3H_6)_9$-$(C_5H_{10})_9$-$C_5H_{11}$.

## 3.1  Molecular Mapping

Before utilizing RelRes, the mapping of atoms into groups must be defined. The previous studies suggested that to keep $r_s$ manageable (i.e., around the pairwise distance associated with nearest neighbors), the CG site should be roughly within one bond length from the FG sites.[25-26] Taking this into consideration, let us consider an arbitrary hydrocarbon molecule. Each "united" atom can be bonded to one, two, three, or four other "united" atoms. A couple of these scenarios are depicted in the top portion of Fig. 3; the "two-case" is shown on the left, and the "four-case" is shown on the right. In our protocol for this article, the central "united" atom will always be the hybrid site, while the peripheral "united" atoms will always be the ordinary sites. The mapping ratio of each group is equal to the total number of "united" atoms it contains.

Of course, such groups, if bonded between each other, can be the building blocks for various oligomeric and polymeric chains. Hence, we have the nomenclature for our work: While a building block by itself is a monomer, two building blocks combined together form a dimer. In turn, several possibilities are presented in the bottom portion of Fig. 3. Let us consider the subtle ramifications associated with these: A group consisting of three sites ($n = 3$, consistently on the left of Fig. 3) can be $C_3H_8$ as a standalone molecule, $C_3H_7$ if at the edge of a chain, or $C_3H_6$ if in the middle of a chain; a group consisting of five sites ($n = 5$, consistently on the right of Fig. 3) can be $C_5H_{12}$ as a standalone molecule, $C_5H_{11}$ if at the edge of a chain, or $C_5H_{10}$ if in the middle of a chain.

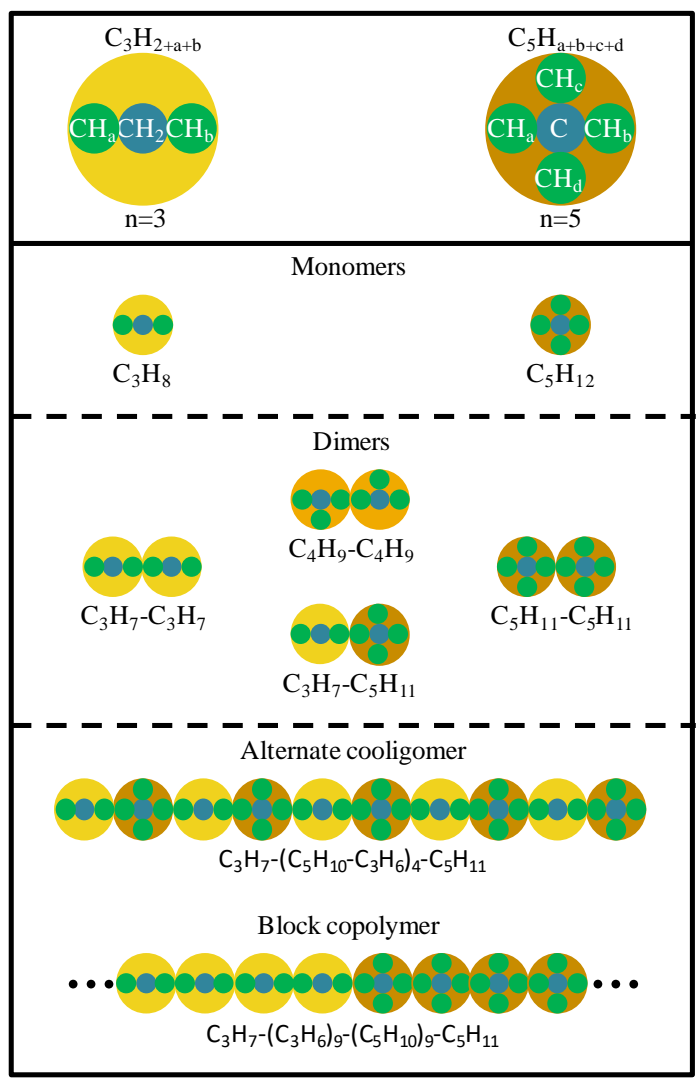

FIG. 3. The mapping of alkane molecules. The orange circles delineate an effective boundary of a group; light orange is for $n = 3$, and dark orange is for $n = 5$ (the former is always on the left, and the latter is always on the right). Actual atoms are given in green: Hybrid sites are represented by dark green circles, while ordinary sites are represented by light green circles. The top portion focuses on two examples of our building blocks for any alkane chain. Subscripts "a", "b", "c" and "d" represent the number of hydrogens in a united atom. The bottom portion, with its three sections, presents the various molecules in our study. Each monomer consists just of one building block ($C_3H_8$ or $C_5H_{12}$). Each dimer consists of two building blocks: Some of the dimers are just symmetric combinations of our two monomers ($C_3H_7$-$C_3H_7$ or $C_5H_{11}$-$C_5H_{11}$); another dimer is an asymmetric combination of our two monomers ($C_3H_7$-$C_5H_{11}$) having a characteristic mapping ratio of 4, while another dimer is a symmetric combination of a different monomer ($C_4H_9$-$C_4H_9$) with $n = 4$. There are also two chain molecules with a characteristic mapping ratio of 4: an alternate cooligomer ($C_3H_7$-($C_5H_{10}$-$C_3H_6$)$_4$-$C_5H_{11}$) and a block copolymer ($C_3H_7$-($C_3H_6$)$_9$-($C_5H_{10}$)$_9$-$C_5H_{11}$); the dots indicate the repeating of groups.

From the perspective of LAMMPS, we must make an important clarification regarding the definitions of atom types in the context of such mapping. As an illustration, we present here an excerpt of the data file of LAMMPS for the dimer $C_3H_7$-$C_5H_{11}$ (shown in the middle section of the bottom portion of Fig. 3):

```
Atoms
   1     1     1    # Hybrid     CH2    C3H7
   2     1     3    # Ordinary   CH2    C3H7
   3     1     5    # Ordinary   CH3    C3H7
   4     1     2    # Hybrid     C      C5H11
   5     1     4    # Ordinary   CH2    C5H11
   6     1     6    # Ordinary   CH3    C5H11
   7     1     6    # Ordinary   CH3    C5H11
   8     1     6    # Ordinary   CH3    C5H11
   9     1     1    # Hybrid     CH2    C3H7
  10     1     3    # Ordinary   CH2    C3H7
  11     1     5    # Ordinary   CH3    C3H7
  12     1     2    # Hybrid     C      C5H11
  13     1     4    # Ordinary   CH2    C5H11
  14     1     6    # Ordinary   CH3    C5H11
  15     1     6    # Ordinary   CH3    C5H11
  16     1     6    # Ordinary   CH3    C5H11
```

We notably focus on the `Atoms` section of the data file, showing just two molecules. Each line corresponds to a particular site, with the position omitted for clarity. The first column specifies the identifier of the site, while the second column specifies the identifier of its molecule. Most importantly, column three defines the atom type of the site, and the comment following the pound sign uniquely characterizes this atom type: It is given by its category (ordinary or hybrid), its atom (C, $CH_2$, or $CH_3$), and its group ($C_3H_7$ or $C_5H_{11}$). Following all possible combinations for the dimer, there are six atom types in total, specifically three atom types for each group (two correspond with ordinary sites, and one corresponds with a hybrid site). Do realize that here, $CH_2$ as an ordinary site and $CH_2$ as a hybrid site have same FG parameters yet different CG parameters, and therefore, they must be defined as distinct atom types.

On another note, this asymmetric dimer brings further subtleties for RelRes: The mapping ratio is different for its two groups. In this case, it is convenient to introduce the concept of a characteristic mapping ratio for the entire system. We define it as the ratio between the number of all sites and the number of hybrid sites. With this definition, observe that for $C_3H_7$-$C_5H_{11}$ the characteristic mapping ratio is 4 (the building blocks themselves have $n = 3$ and $n = 5$). Note that the oligomeric-polymeric solutions also have 4 as the characteristic mapping ratio.

### 3.2 Setup of Molecular Simulations

Without loss of generality, for the non-bonded interactions of the reference models, we use here the set of LJ parameters of OPLS_UA.[47] These parameters are presented in Table 1 (in the FG column). For the bonded interactions, we utilize the AMBER_UA parameters, which are commonly used in conjunction with OPLS_UA[48]: We use a harmonic potential for the bond stretching (with a bond length of 0.1526nm and a spring constant of 180,640zJ/nm$^2$), a harmonic potential for the bond bending (with a bond angle of 1.9618rad and a spring constant of 437.7zJ/rad$^2$), and a Fourier potential calculated as $13.9\text{zJ}(1 + \cos 3\gamma)$ for a dihedral angle $\gamma$, concurrently scaling down the "1-4" non-bonded interactions by a factor of 2.0.

For the RelRes systems, two further aspects of the non-bonded interactions are required: The LJ energy and length parameters of the CG potential, together with the switching distance, must be defined. In general, once groups with their reference FG parameters are defined, the respective CG parameters are calculated using Eq. (15) (since geometric mixing is applicable for OPLS_UA). These are presented in Table 1 (in the CG column). In selecting $r_s$, we consider the outcome from Refs. 25-26: The switching distance should be between the first and second coordination shells. We essentially satisfy this condition for all our systems, having our switching distances vary from 0.6nm to 0.7nm; as for the smoothing zone, we use $\Delta r_s = 0.05$nm throughout. Besides in all molecular simulations, we employ an inner $r_{c-} = 1.2$nm and an outer $r_{c+} = 1.4$nm. In line with previous studies,[25-26] for the bonded energetics in the RelRes systems, we use exactly the same parameters as we do in the reference models.

We employ a box with periodic boundaries for all molecular simulations. Our monomeric systems consist of $N = 2000$ molecules, and our dimeric systems consist of $N = 1000$ molecules. In the case of the oligomeric-polymeric solutions, there are $N = 1920$ solvent molecules, together with four cooligomers or two copolymers; in both cases, the mass fraction of the chains is about 0.02. The following temperature and pressure are established throughout our study: $T = 4.0$zJ (290 K) and $P = 2.0$MJ/m$^3$ (15,000 Torr): This ensures that all systems are in the liquid phase.

To achieve the desired state condition, a preliminary run of each simple system (monomers and dimers) was executed for 4.0ns (with a timestep of 4.0fs) under the isothermal-isobaric ensemble using the reference models (pair style lj/smooth in LAMMPS). After establishing the desired state condition, the barostat was removed yet the thermostat was maintained, meaning that all molecular simulations were formally conducted in the conventional canonical ensemble. For each RelRes model (with its unique set of parameters), as well as for the reference model itself, we used the following protocol: We performed the equilibration for 1.0ns (with a timestep of 2.0fs) and the production for 1.0ns (with a timestep of 1.0fs). The protocol for the oligomeric-polymeric solutions was somewhat different, since the chains move slowly. Foremost, the preliminary run (with the isothermal-isobaric ensemble) was done with the RelRes model rather than with the reference model. Then, once in the canonical ensemble, we performed the equilibration for 4.0ns (with a timestep of 2.0fs) and the production for 10.0ns (with a timestep of 1.0fs).

To compare the structural behavior of the fluids, we notably computed the radial distributions $g(r)$ for all combinations of atom types; these were sampled every 10 timesteps, while averaging over 50,000 samples. For the oligomeric-polymeric solutions, $g(r)$ was generated only for the solvent molecules, while for the solute chains, their radius of gyration $R_g$ was calculated instead, based on data collected every 200 timesteps. Besides, the displacement of all molecules was recorded every 200 timesteps, and in turn, the diffusion coefficient $D$ was calculated from the slope of the squared displacement.

Table 1. Non-bonded interactions parameters. The top portion of the table addresses ordinary sites (these just involve FG parameters, with CG parameters being not applicable); the bottom portion addresses hybrid sites. For the latter, the parametrization between FG and CG parameters is done by Eq. (15).

| Type | Atom | Group | FG $\epsilon^{FG}$ [zJ] | FG $\sigma^{FG}$ [nm] | CG $\epsilon^{CG}$ [zJ] | CG $\sigma^{CG}$ [nm] |
|---|---|---|---|---|---|---|
| Ordinary | CH$_3$ | C$_3$H$_8$ C$_3$H$_7$ | 1.2158 | 0.3905 | 0.0 | 0.0 |
| | | C$_4$H$_{10}$ C$_4$H$_9$ | 1.1116 | 0.3910 | | |
| | | C$_5$H$_{12}$ C$_5$H$_{11}$ C$_5$H$_{10}$ | 1.0074 | 0.3960 | | |
| | CH$_2$ | C$_3$H$_7$ C$_3$H$_6$ C$_4$H$_9$ C$_5$H$_{11}$ C$_5$H$_{10}$ | 0.8198 | 0.3905 | 0.0 | 0.0 |
| Hybrid | CH$_2$ | C$_3$H$_8$ | 0.8198 | 0.3905 | 9.6768 | 0.3905 |
| | | C$_3$H$_7$ | | | 8.4887 | 0.3905 |
| | | C$_3$H$_6$ | | | 7.3785 | 0.3905 |
| | CH | C$_4$H$_{10}$ | 0.5558 | 0.3850 | 15.267 | 0.3899 |
| | | C$_4$H$_9$ | | | 14.126 | 0.3897 |
| | C | C$_5$H$_{12}$ | 0.3474 | 0.3800 | 21.133 | 0.3942 |
| | | C$_5$H$_{11}$ | | | 20.237 | 0.3931 |
| | | C$_5$H$_{10}$ | | | 19.366 | 0.3919 |

To compare the thermal behavior of the fluids, we calculated intra-molecular $U_\psi$ and inter-molecular $\widetilde{U}$ potential energy. In addition, we calculated the virial associated with intra-molecular $W_\psi$ and inter-molecular $\widetilde{W}$ forces. For the oligomeric-polymeric solutions, the energy was calculated separately for the solvent and solute molecules. All relevant data was outputted every 200 timesteps. This data was used for calculating static properties such as arithmetic means (e.g., $\langle U_\psi \rangle$, $\langle \widetilde{U} \rangle$, etc.) and standard deviations (e.g., $\langle \delta U_\psi^2 \rangle^{1/2}$, $\langle \delta \widetilde{U}^2 \rangle^{1/2}$, etc.). This data was also used for evaluating dynamic properties, such as auto-correlations in terms of time $t$ (e.g., $\langle \delta W_{\psi t} \delta W_{\psi 0} \rangle / \langle \delta W_\psi^2 \rangle$, $\langle \delta \widetilde{W}_t \delta \widetilde{W}_0 \rangle / \langle \delta \widetilde{W}^2 \rangle$, etc.). Note that we frequently normalize thermal properties, denoted by the symbol "*": This corresponds with a ratio in a property between the RelRes system and the reference system.

### 3.3 Elementary Monomers

We begin our investigation with the two monomeric liquids, each one having its own mapping ratio (i.e., $C_3H_8$ with $n = 3$ and $C_5H_{12}$ with $n = 5$; see mapping in the upper section of the bottom portion of Fig. 3). Our main goal here is in determining the ideal switching distance for a given mapping ratio. Besides, we also validate here the parametrization of Eq. (15). Consequently, the monomers become the basis for all other systems we test.

As discussed earlier, a shorter $r_s$ corresponds with a faster computation, and thus, the ideal $r_s$ is given as the shortest switching distance that allows for the correct retrieval of structural and thermal behavior. In the previous work, a conjecture was made that the proper switching distance is between the first and second coordination shells of a certain liquid.[25] It was in turn shown that this specifically corresponds with the midpoint between the maximum and minimum of $g(r)$ (between the centers of mass), which is roughly in the vicinity of the respective inflection point.[26]

For the reference liquids, Fig. 4 presents the radial distribution of the monomers as black solid lines ($C_3H_8$ is shown on the left and $C_5H_{12}$ on the right): The top panels are for the hybrid-hybrid $g(r)$, the bottom panels are for the ordinary-ordinary $g(r)$, and the middle panels are for the hybrid-ordinary $g(r)$. Since the hybrid site is closest to the center of mass, we specifically calculate the signature distances (i.e., the midpoint as well as the inflection point) associated with its radial distribution, and these are presented in Table 2. They lie in the range between

~0.6nm and ~0.7nm. Therefore, we decided to study RelRes systems with two different values of $r_s$ (corresponding with the bounds of this range). This can in turn help us ultimately decide on the most proper value of the switching distance.

We also show in Fig. 4 the comparison of the reference $g(r)$ with the RelRes $g(r)$ for the monomer systems. The blue dashed curves are for $r_s = 0.6$nm, and the red dashed curves are for $r_s = 0.7$nm. Notice that the black lines of the reference model perfectly match all the dashed lines of the RelRes model in the middle and bottom panels (i.e., for $g(r)$ that involve ordinary sites). Only in the top panels, that depict the hybrid-hybrid $g(r)$, there is a noticeable deviation of the RelRes curves from the reference curves. The associated coordinates of the maxima and minima of these curves are presented in Table 3. For $C_3H_8$, although $r_s = 0.7$nm shows a slightly better match, both switching distances give an almost perfect description of the radial distribution of the reference model: The locations and values of the extrema are off by less than 5% for $r_s = 0.6$nm (we thus recommend the latter switching distance for the mapping ratio of 3). Contrarily, for $C_5H_{12}$ with $r_s = 0.6$nm, $g(r_{max})$ is off by about 20%. This is why, from a structural perspective, our recommendation for the proper switching distance for the mapping ratio of 5 is $r_s = 0.7$nm (its $g(r_{max})$ is off by less than 1%).

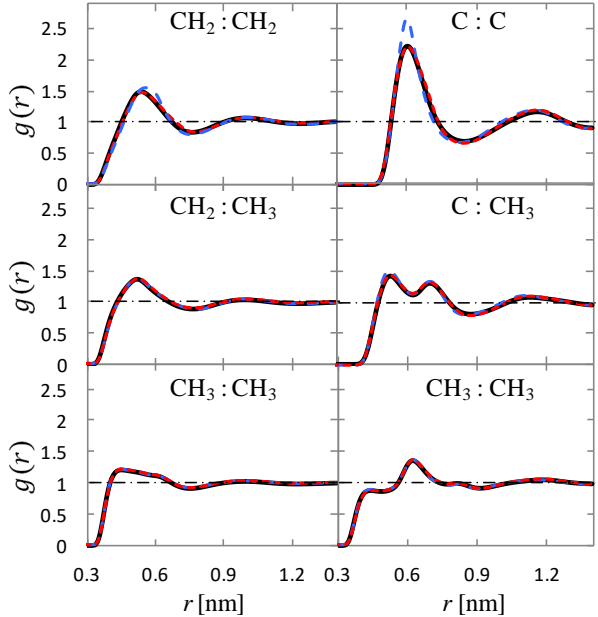

FIG. 4. Radial distribution function for $C_3H_8$ (left set of panels) and $C_5H_{12}$ (right set of panels). The reference system is shown in black solid lines, while RelRes is given in colored dashed lines; the blue is for $r_s = 0.6$nm, and the red is for $r_s = 0.7$nm. The top panels show $g(r)$ between two hybrid sites, the bottom panels show $g(r)$ between two ordinary sites, and the middle panels show $g(r)$ between a hybrid site and an ordinary site.

Table 2. Signature distances in the reference $g(r)$ of all studied liquids. These are specifically the midpoint between the two extrema, as well as its adjacent inflection point. The rightmost column also shows our ultimate suggestions for the switching distances in RelRes.

| Atom | Group | System | $g(r)$ midpoint [nm] | $g(r)$ inflection point [nm] | Suggested $r_s$ [nm] |
|---|---|---|---|---|---|
| CH$_2$ | C$_3$H$_8$ | C$_3$H$_8$ | 0.65 | 0.66 | |
| | C$_3$H$_7$ | C$_3$H$_7$-C$_3$H$_7$ | 0.62 | 0.59 | 0.60 |
| | | C$_3$H$_7$-C$_5$H$_{11}$ | 0.62 | 0.58 | |
| CH | C$_4$H$_{10}$ | C$_4$H$_{10}$ | 0.69 | 0.65 | 0.65 |
| | C$_4$H$_9$ | C$_4$H$_9$-C$_4$H$_9$ | 0.67 | 0.64 | |
| C | C$_5$H$_{12}$ | C$_5$H$_{12}$ | 0.72 | 0.67 | |
| | C$_5$H$_{11}$ | C$_5$H$_{11}$-C$_5$H$_{11}$ | 0.69 | 0.65 | 0.70 |
| | | C$_3$H$_7$-C$_5$H$_{11}$ | 0.70 | 0.65 | |

Table 3. Comparison of structural characteristics for the studied liquids. The coordinates of the minima and maxima for the radial distributions, as well as the diffusion coefficients, are presented for the monomers, dimers, and the solvent of the cooligomer and copolymer systems. For the cooligomeric and copolymeric solutes themselves, the radius of gyration is presented instead. For the systems comprised from different building blocks, characteristic values (marked with *) are given for their mapping ratios, as well as for their switching distances. For $C_3H_7$-$C_5H_{11}$, results for $g(r)$ between different hybrid sites (i.e., $CH_2$:C of the middle panel of Fig. 6) are provided; the intra-molecular extrema of Fig. 6 are excluded.

| | System | $n$ | $r_s$ [nm] | Maxima ($r_{max}$, $g(r_{max})$) | | Minima ($r_{min}$, $g(r_{min})$) | | $R_{\bar{g}}$ [nm] | | $D$ [nm²/ns] | |
|---|---|---|---|---|---|---|---|---|---|---|---|
| | | | | Ref. | RelRes | Ref. | RelRes | Ref. | RelRes | Ref. | RelRes |
| Monomers | $C_3H_8$ | 3 | 0.6 | (0.54, 1.49) | (0.56, 1.56) | (0.76, 0.84) | (0.76, 0.80) | na | | 0.950 | 0.903 |
| | | | 0.7 | | (0.54, 1.47) | | (0.78, 0.83) | | | | 0.916 |
| | $C_5H_{12}$ | 5 | 0.6 | (0.60, 2.22) | (0.60, 2.65) | (0.85, 0.69) | (0.83, 0.66) | | | 0.392 | 0.366 |
| | | | 0.7 | | (0.61, 2.20) | | (0.84, 065) | | | | 0.365 |
| Dimers | $C_3H_7$-$C_3H_7$ | 3 | 0.6 | (0.52, 1.35) | (0.53, 1.36) | (0.73, 0.81) | (0.73, 0.78) | na | | 0.273 | 0.263 |
| | $C_5H_{11}$-$C_5H_{11}$ | 5 | 0.7 | (0.60, 2.02) | (0.60, 2.00) | (0.79, 0.68) | (0.79, 0.65) | | | 0.070 | 0.067 |
| | $C_3H_7$-$C_5H_{11}$ | *4* | *0.65* | (0.55, 1.58) | (0.56, 1.56) | (0.76, 0.73) | (0.76, 0.70) | | | 0.135 | 0.135 |
| | $C_4H_9$-$C_4H_9$ | 4 | 0.65 | (0.57, 1.53) | (0.58, 1.54) | (0.76, 0.76) | (0.76, 0.73) | | | 0.162 | 0.154 |
| Complex Solutions | Cooligomer | *4* | *0.65* | (0.58, 1.71) | (0.59, 1.76) | (0.80, 0.77) | (0.80, 0.73) | 0.77 | 0.78 | 0.619 | 0.567 |
| | Copolymer | *4* | *0.65* | (0.58, 1.72) | (0.59, 1.76) | (0.80, 0.77) | (0.80, 0.73) | 1.28 | 1.29 | 0.607 | 0.554 |

Table 3 also presents the diffusion coefficient. RelRes adequately represents this facet of the reference system, having less than 10% discrepancy for it, which is within its statistical error. This good representation stems mostly in the fact that RelRes keeps all degrees of freedom. Moreover, as the diffusivity is well correlated with many other transport coefficients (e.g., viscosity), we expect that RelRes will yield an adequate representation for these as well.

Besides these structural correlations, we find that RelRes also captures thermal properties satisfactorily well. Table 4 shows the potential and virial associated with the inter-molecular energetics, specifically presenting the values of their corresponding arithmetic means and standard deviations. In all cases, we naturally find the overall trend that the larger $r_s$ yields a smaller discrepancy in both the potential and virial energies. For $C_3H_8$, we find sufficiently good results with $r_s = 0.6$nm: The discrepancy in the mean values is always less than 1%, while the discrepancy in the deviation values is always less than 7%. For $C_5H_{12}$, we find sufficiently good results only with $r_s = 0.7$nm: The discrepancy in the mean values is always less than 3%, while the discrepancy in the deviation values is always less than 3%. These discrepancies are more or less within the range of the statistical errors. Note that we also look at the intra-molecular energetics of the potential and virial (not presented in Table 4): In all cases, we observe no noticeable discrepancies, and this is expected, since our multiscale strategy does not alter this aspect of the reference liquid. This concludes our recommendation for the proper switching distance: $r_s = 0.6$nm for the mapping ratio of $n = 3$, and $r_s = 0.7$nm for the mapping ratio of $n = 5$. Overall, this again nicely corresponds with the signature distances we emphasized in Table 2.

Table 4. Comparison of thermal characteristics for the studied liquids. Specifically, the arithmetic means and standard deviations are given for the pairwise potential and associated virial. In all cases, the properties are normalized by the total number of molecules in the system. For the systems comprised from different building blocks, characteristic values (marked with *) are given for their mapping ratios, as well as for their switching distances.

| | System | $n$ | $r_s$ [nm] | $\langle \widetilde{U} \rangle / N$ [zJ] Ref. | $\langle \widetilde{U} \rangle / N$ [zJ] RelRes | $\langle \delta \widetilde{U}^2 \rangle^{1/2} / N$ [zJ] Ref. | $\langle \delta \widetilde{U}^2 \rangle^{1/2} / N$ [zJ] RelRes | $\langle \widetilde{W} \rangle / N$ [zJ] Ref. | $\langle \widetilde{W} \rangle / N$ [zJ] RelRes | $\langle \delta \widetilde{W}^2 \rangle^{1/2} / N$ [zJ] Ref. | $\langle \delta \widetilde{W}^2 \rangle^{1/2} / N$ [zJ] RelRes |
|---|---|---|---|---|---|---|---|---|---|---|---|
| Monomers | $C_3H_8$ | 3 | 0.6 | -19.3 | -19.5 | 0.076 | 0.081 | -8.39 | -8.35 | 0.472 | 0.472 |
| Monomers | $C_3H_8$ | 3 | 0.7 | -19.3 | -19.3 | 0.076 | 0.078 | -8.39 | -8.14 | 0.472 | 0.451 |
| Monomers | $C_5H_{12}$ | 5 | 0.6 | -28.9 | -30.6 | 0.085 | 0.097 | -15.28 | -19.04 | 0.530 | 0.603 |
| Monomers | $C_5H_{12}$ | 5 | 0.7 | -28.9 | -29.5 | 0.085 | 0.086 | -15.28 | -14.95 | 0.530 | 0.524 |
| Dimers | $C_3H_7$-$C_3H_7$ | 3 | 0.6 | -48.3 | -48.1 | 0.180 | 0.186 | -6.88 | -6.05 | 1.048 | 1.058 |
| Dimers | $C_5H_{11}$-$C_5H_{11}$ | 5 | 0.7 | -19.9 | -19.6 | 0.390 | 0.392 | 211.43 | 213.16 | 1.873 | 1.789 |
| Dimers | $C_3H_7$-$C_5H_{11}$ | *4* | *0.65* | -34.0 | -33.8 | 0.304 | 0.303 | 104.27 | 105.90 | 1.448 | 1.525 |
| Dimers | $C_4H_9$-$C_4H_9$ | 4 | 0.65 | -40.0 | -39.8 | 0.297 | 0.299 | 85.25 | 85.93 | 1.438 | 1.478 |
| Complex Solutions | Cooligomer | *4* | *0.65* | -25.0 | -25.4 | 0.091 | 0.092 | -10.19 | -10.31 | 1.570 | 1.495 |
| Complex Solutions | Copolymer | *4* | *0.65* | -25.2 | -25.6 | 0.090 | 0.094 | -9.95 | -10.06 | 1.583 | 1.621 |

For these monomers, we conducted an additional study to verify that the parametrization from the FG potential to the CG potential, as defined by Eq. (15), is in fact optimal. For this purpose, we introduce two tuning factors, $\lambda_\epsilon$ and $\lambda_\sigma$, and we vary them while running the RelRes simulations with CG parameters $\lambda_\epsilon \epsilon^{CG}$ and $\lambda_\sigma \sigma^{CG}$ (note that $\lambda_\epsilon = 1.0$ and $\lambda_\sigma = 1.0$ yield the ideal parameters of Eq. (15)). In Fig. 5, we plot several thermal properties as functions of $\lambda_\epsilon$ (bottom abscissa, squares, darker solid lines) and $\lambda_\sigma$ (top abscissa, diamonds, lighter dashed lines): Specifically, we plot the inter-molecular potentials in the top panels, and the intra-molecular potentials in the bottom panels; the arithmetic mean is in red, and the standard deviation is in green. Again, the left panels are for $C_3H_8$, and the right panels are for $C_5H_{12}$. Error bars are not shown: For the energy mean, they are within the size of the markers, and for the energy deviation, they are in the range of 5%. Bottom panels confirm that the intra-molecular energy is not sensitive to the CG parameters of RelRes.

It is interesting to notice in the top panels that there is linear dependency of the inter-molecular energy with both tuning factors. The scales of the top and bottom abscissa were specifically chosen to be different such that $\frac{\lambda_\epsilon}{\lambda_\sigma^6} \approx 1$. Observe now that with these scales, the slopes of the curves are almost identical, which implies that the value of the sixth power term $(-4\epsilon^{CG}(\sigma^{CG}/r)^6)$ of the LJ potential has the most significant influence on the thermal behavior. This suggests that the parameters which are derived analytically by Eq. (15) can be perhaps better optimized by varying $\lambda_\epsilon$ and $\lambda_\sigma$ such that $\lambda_\epsilon \lambda_\sigma^6 = 1$. Use of the relative entropy for such optimization may be helpful,[45-46] but it will obviously require significant computational resources as compared with the analytical relation of Eq. (15), and thus, it is beyond the scope of our current work.

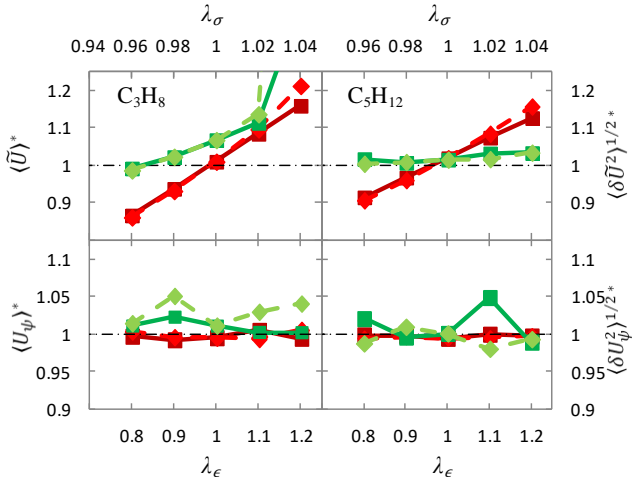

FIG. 5. The components of the potential energy, normalized by their respective values from the reference system. The top panels are for the inter-molecular energy, and the bottom panels are for the intra-molecular energy; the left panels are for $C_3H_8$, and the right panels are for $C_5H_{12}$. The data is plotted in terms of $\lambda_\epsilon$ (squares connected by solid lines, plotted on the bottom abscissa) and $\lambda_\sigma$ (diamonds connected by dashed lines, plotted on the top abscissa). The red color is for the arithmetic mean of the energies (plotted on the left ordinate), and the green color is for the standard deviation of the energies (plotted on the right ordinate).

### 3.4 Various Dimers

Proceeding with RelRes for dimers, we adopt the recommendations from the analysis of the monomers: $r_s = 0.6$nm for $n = 3$ ($C_3H_7$) and $r_s = 0.7$nm for $n = 5$ ($C_5H_{11}$). In turn, we run simulations with the same protocol as for the monomers. Our main aim here is in determining if the switching distances are mostly governed by the mapping ratio (rather than by any peculiarities in the molecular complexities).

Let us discuss briefly the results for the two symmetric dimers (i.e., $C_3H_7$-$C_3H_7$ and $C_5H_{11}$-$C_5H_{11}$, see the mapping for these dimers in the middle section of the bottom portion of Fig. 3).

For these dimers, the replication capability of both structural and thermal behavior is very similar to that of the monomers. Detailed results are presented in Tables 3 and 4. One can notice that the replication errors for the dimers are even smaller than for the monomers. For example, for $C_3H_7$-$C_3H_7$, $r_{max}$ and $g(r_{max})$ are off by about 2% and 1% respectively, while the corresponding discrepancies for $C_3H_8$ are 3% and 5% respectively. Looking into Table 2, we observe that both signature distances (i.e., the midpoint with the inflection point) for the dimers is lower than for the monomers. This fact suggests that as compared with the monomers, the dimers may allow lower $r_s$ or perform better with the same $r_s$. In any case, for consistency, we recommend to use the same switching distance for all groups with the same mapping ratio.

The asymmetric dimer $C_3H_7$-$C_5H_{11}$ needs to be discussed separately (see the mapping in the middle section of the bottom portion of Fig. 3). It was especially included in our study in order to examine RelRes on a molecule that has different groups. As we recommend $r_s = 0.6$nm for $n = 3$ and $r_s = 0.7$nm for $n = 5$, this dimer system notably entails different switching distances. How about the value of the switching distance for the mixed interaction between the different groups? In this case, geometric mixing for $r_s$ is applied: $r_s = \sqrt{0.6 \cdot 0.7}$nm $\cong 0.6481$nm $\cong 0.65$nm. This value can be considered as a characteristic switching distance for the system, since it represents the average of all individual switching distances involved. The study of this asymmetric dimer provides a baseline for the study of more complex systems, which possess significantly more asymmetry.

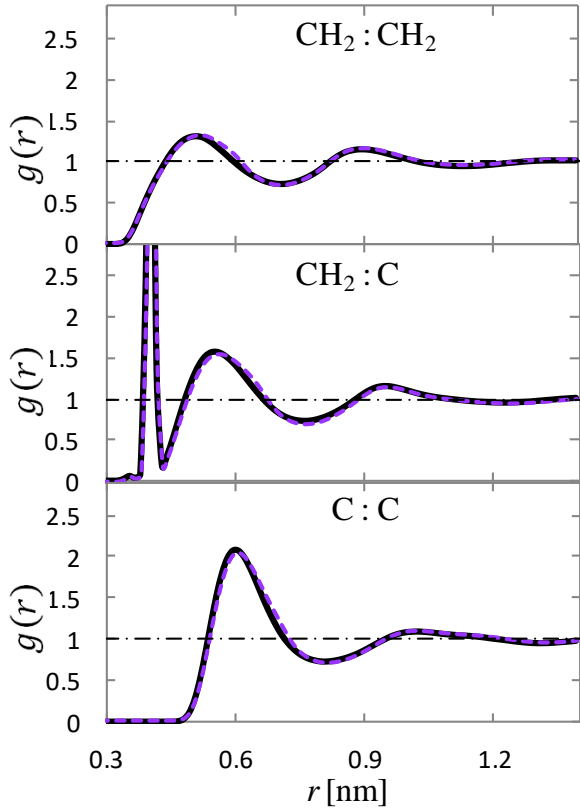

FIG. 6. Radial distribution function for $C_3H_7$-$C_5H_{11}$. The reference system is shown in black solid lines, while RelRes is given in violet dashed lines. The top panel shows $g(r)$ between hybrid sites $CH_2$ (from the $C_3H_7$ group), the bottom panel shows $g(r)$ between hybrid sites C (from the $C_5H_{11}$ group), and the middle panel shows $g(r)$ between hybrid sites $CH_2$ and C (from different groups, $C_3H_7$ and $C_5H_{11}$ respectively). The extrema around $r = 0.4$nm in the middle panel is attributed to the intra-molecular pair.

Recall that $C_3H_7$-$C_5H_{11}$ has two distinct hybrid sites and four distinct ordinary sites. We analyzed $g(r)$ for all combinations of sites and have found that similarly as with the monomers and the other dimers, ordinary-ordinary $g(r)$ and hybrid-ordinary $g(r)$ of RelRes perfectly match reference $g(r)$. Only the three hybrid-hybrid $g(r)$ are slightly impacted by RelRes. They are shown in Fig. 6. Reference $g(r)$ is shown in black solid lines, and RelRes $g(r)$ is shown in violet dashed lines. The top panel presents the pure $g(r)$ for hybrid sites of groups with a mapping ratio of 3, the bottom panel shows the pure $g(r)$ for hybrid sites of groups with a mapping ratio of 5, and finally, the middle panel shows the mixed $g(r)$ between hybrid sites of groups with different mapping ratios. We can see on all panels that RelRes adequately represents structural behavior of the reference model. We can see the same in Tables 3 and 4: The data confirms that the recommended switching distances provide a good representation of the arithmetic mean and standard deviation of the energy, the virial, as well as the diffusion coefficient. Fig. 7 shows a comparison of a dynamic correlation of two thermal functions, intra-molecular and inter-molecular components of the virial. It also shows a good match between the RelRes and reference models.

The fourth dimer, the symmetric dimer $C_4H_9$-$C_4H_9$, was selected to participate in the study because its groups have a mapping ratio of $n = 4$. Initially, we did not study the associated monomer $C_4H_{10}$. What should be the proper switching distance for this dimer? The intuitive approach is a basic interpolation between the other two mapping ratios; this yields $r_s = 0.65$nm for $n = 4$. To confirm this, let us look at $g(r)$ of the reference liquid (Table 2). The midpoint is 0.67nm and the inflection point is 0.64nm. So, a choice of $r_s = 0.65$nm does seem ideal. Results for molecular simulations using this $r_s$ are presented in Tables 3 and 4 for structural and thermal comparison.

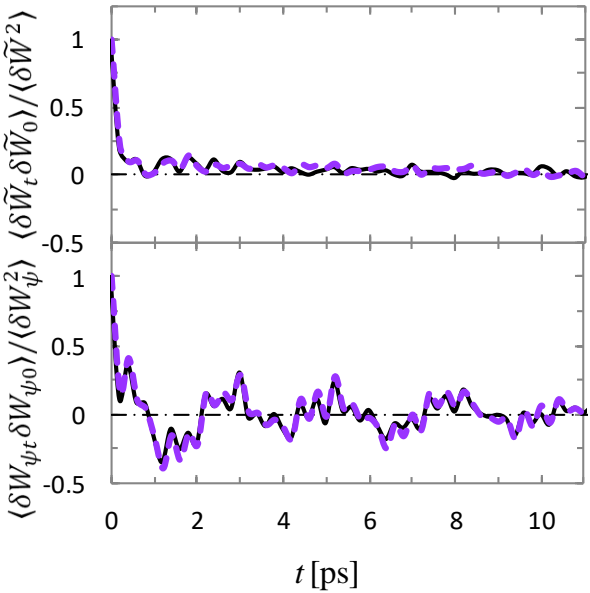

FIG. 7. Dynamic correlations of the virial for $C_3H_7$-$C_5H_{11}$. The top panel shows its inter-molecular component, and the bottom panel shows its intra-molecular component. The reference system is shown with solid black lines, while RelRes is given with dashed violet lines.

Graphs are not shown as they are very similar to all other liquids studied.

The study of the dimers confirms the robustness of our multiscale strategy. Foremost, the selection of the switching distance is mainly dependent on the mapping ratio of the group. Interestingly, with an increase of size of molecules (i.e., from monomers to dimers, etc.), the performance of RelRes even slightly improves with the same switching distance. This gives a good hope that this improvement will be even better for more complex systems.

### 3.5 Complex Solutions

In the last stage of our verification, we dilute a few cooligomers or copolymers in a solvent, building in turn two complex solutions. For the solvent, we use isobutane ($C_4H_{10}$) in both systems. In one system, we immerse an alternate cooligomer $C_3H_7$-$(C_5H_{10}$-$C_3H_6)_4$-$C_5H_{11}$. This cooligomer has 10 groups: one $C_3H_7$ group, four $C_3H_6$ groups, four $C_5H_{10}$ groups and one $C_5H_{11}$ group. In the second system, we immerse a block copolymer $C_3H_7$-$(C_3H_6)_9$-$(C_5H_{10})_9$-$C_5H_{11}$. This copolymer has 20 groups: one $C_3H_7$ group, nine $C_3H_6$ groups, nine $C_5H_{10}$ groups and one $C_5H_{11}$ group. The mapping of these two systems is shown in the lower section of the bottom portion of Fig. 3.

For the complex solutions, the switching distances were based on the mapping ratio of each group: Relevant for the solutes, $r_s = 0.6$nm for the $C_3H_7$ and $C_3H_6$ groups ($n = 3$), and $r_s = 0.7$nm for the $C_5H_{11}$ and $C_5H_{10}$ groups ($n = 5$); regarding the solvent, $r_s = 0.65$nm for the $C_4H_{10}$ group ($n = 4$). Similar to the dimer $C_3H_7$-$C_5H_{11}$, the value of $r_s \cong 0.65$nm represents a characteristic switching distance of the systems.

First, let us discuss the behavior of the solute. In both systems, we are getting a good representation of the properties. From a structural perspective (see Table 3), the radius of gyration in both cases is within 1% from the reference system, which is in the range of the statistical error (about 2%). From a thermal perspective, intra-molecular energy is mainly not impacted by RelRes. For inter-molecular energy, the results are also good: Both the mean and deviation in the energy is within the statistical error of roughly 10%. To make these results evident, we produced graphs of probability distribution $\mathcal{P}$ of components of the potential energy, specifically for the copolymer molecules. They are presented in Fig. 8. The top panel is for the inter-molecular energy (upper abscissa), and the bottom panel is for the intra-molecular component (lower abscissa). The reference system is shown as the black solid line, and the RelRes system is shown as the violet dashed line.

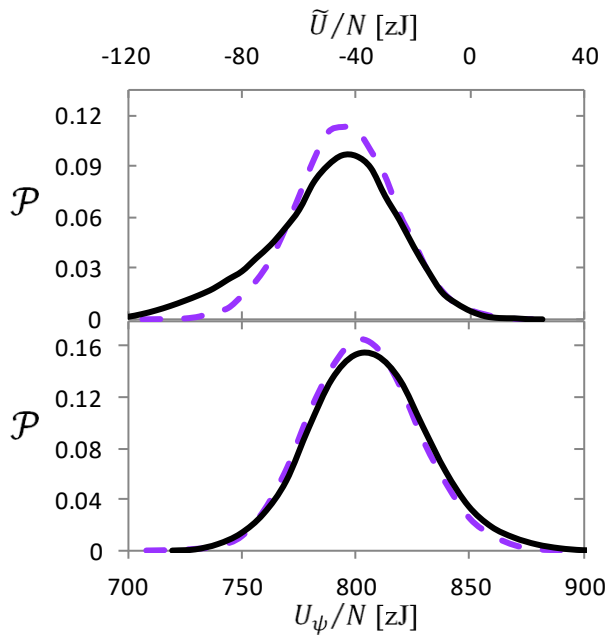
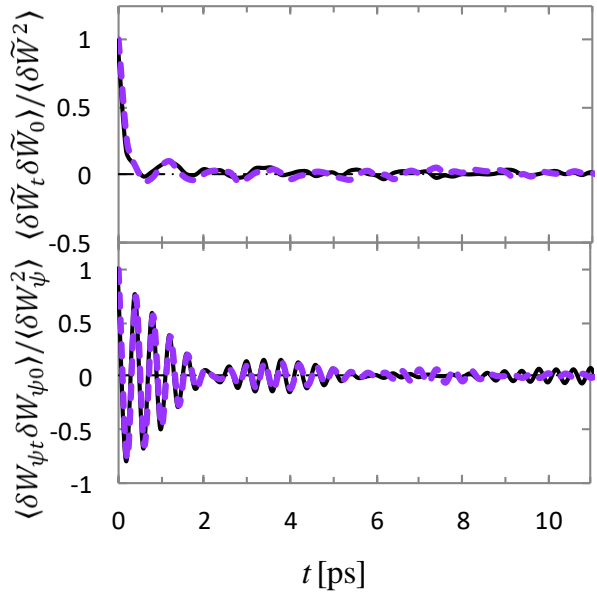

FIG. 8. Probability distribution for the potential energy of the copolymer. The top panel is for the inter-molecular component (plotted on the top abscissa), and the bottom panel is for the intra-molecular component (plotted on the bottom abscissa). Black solid curves depict the reference data, while RelRes is plotted in violet dashed lines.

FIG. 9. Dynamic correlations of the virial for the copolymer solution. The top panel shows its inter-molecular component, and the bottom panel shows its intra-molecular component. The reference system is shown with solid black lines, while RelRes is given with dashed violet lines.

Let us now discuss the solvent. With $r_s = 0.65$nm, in both solutions, RelRes perfectly describes structural (see Table 3) and thermal (not shown) behavior of the reference model, and the properties of the solvent are very similar in both cases. Besides, curves of dynamic correlation of virial components for the copolymer solution are plotted in Fig. 9, which is analogous in format to Fig. 7. It also shows a good match between the models. For cooligomers, graphs are very similar, yet they are not shown here.

### 3.6 Computational Efficiency

At this point, we have shown that RelRes in LAMMPS can capture the behavior of nonpolar liquids very well. However, is it actually worth using in consideration of computational efficiency? Here, we in fact show that this is the case.

A separate test was performed for all systems to analyze the computational efficiency of the RelRes method. Such a test was started from the last configuration of the production run for each of the systems, and it consisted of four separate runs of 0.05ns (with a timestep of 1.0fs). Note that LAMMPS, in its log file, reports the statistics of the timing breakdown: In particular, it has the total time $\tau_{\text{tot}}$, and it has the time taken just by the pairwise interactions $\tilde{\tau}$; we use these CPU times to respectively calculate the computational efficiencies $\eta^*_{\text{tot}}$ and $\tilde{\eta}^*$. Besides, the log file also contains the average size of the neighbor list, which we take as our $\mathbb{N}$. We expect that this value signals the speedup of RelRes as given by Eqs. (21) and (22).

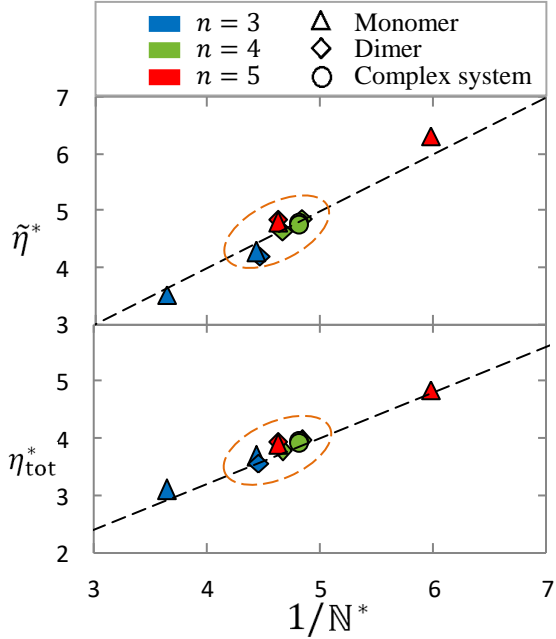

FIG. 10. The computational efficiency relative to the inverse of the characteristic size of the neighbor list. The top panel shows the efficiency associated just with the pairwise interactions, while the bottom panel shows the total efficiency. Different colors represent different mapping ratios: $n = 3$ is plotted in blue, $n = 4$ is plotted in green, and $n = 5$ is plotted in red. Different symbols signify different molecular complexities: Monomers are represented by triangles, dimers are represented by diamonds, and oligomeric-polymeric solutions are represented by circles. The black dashed lines are the expected trends given by Eqs. (21) and (22) using $\chi = 4/5$. The dashed ellipse indicates the region in the graph which encompasses the systems with the recommended switching distances.

Fig. 10 plots the computational efficiencies as a function of the inverse of the list size: The top panel presents the efficiency associated just with the pairwise interactions, and the bottom panel presents the total efficiency. Different colors denote different mapping ratios, while different symbols denote different molecular complexities. Error bars are not shown as they are all negligible.

Focus for now just on the region encompassed by the dashed ellipse. This is the region of all systems with the recommended switching distances, as given by Table 2. Most importantly, the molecular simulations using RelRes are almost fivefold faster than the corresponding reference systems in calculating the pairwise interactions, and in turn, they are about fourfold faster in terms of the total CPU time. Besides, notice that in this region, all symbols of the same color essentially collapse on each other: This means that with the recommended choice for the switching distance, systems with the same mapping ratio have an almost identical computational efficiency, rendering the computational enhancement of RelRes independent of the molecular complexity. Therefore, we expect a similar speedup of almost an order magnitude for other nonpolar mixtures using RelRes.

We now shift our focus on the triangles of Fig. 10, which represent the two monomers, $C_3H_8$ and $C_5H_{12}$. In the elliptical region, the blue triangle is for $n = 3$ with $r_s = 0.6$nm, and the red triangle is for $n = 5$ with $r_s = 0.7$nm. Going beyond the elliptical region, the lowest blue triangle is for $n = 3$ with $r_s = 0.7$nm, and the highest red triangle is for $n = 5$ with $r_s = 0.6$nm (realize that these switching distances are not the recommended ones). Regarding the switching distance, given the same color of triangles, the higher $r_s$ consistently yields the lower computational efficiency. Regarding the mapping ratio, the higher $n$ consistently renders the higher computational efficiency. Altogether, this confirms our earlier intuition that decreasing $r_s$ and increasing $n$ are the main factors for improving the computational efficiency of RelRes.

Finally, focus on the black dashed lines in the top and bottom panels of Fig. 10. These respectively confirm our predictions formulated in Eqs. (21) and (22): The computational efficiency is directly proportional to the reduction in the characteristic size of the neighbor list. Besides, we also perform a linear regression through the origin (not shown in Fig. 10): For $\tilde{\eta}^*$, we find a slope of 1.0065 with a determination coefficient of 0.94, and for $\eta^*_{tot}$, we find a slope of 0.8248 with a determination coefficient of 0.96. These trends affirm that RelRes can reduce the CPU time of reference systems by almost an order of magnitude.

## 4  Conclusion

Here, we have continued the work on RelRes, the multiscale algorithm which describes near neighbors with FG models and far neighbors with CG models.[23-26] Foremost, we automated RelRes in the computational package of LAMMPS, particularly for the LJ potential. Indeed, we developed a new C++ class, `pair_lj_relres`, with its pair style, `lj/relres`, that calculates pairwise interactions according to Eq. (20). Most importantly in the current work, we demonstrated that RelRes is in fact computationally efficient. In particular, our implementation in LAMMPS reduces the computational time of molecular simulations by a factor of about 4-5.

We have essentially tested this implementation on several alkane liquids, being as complex as solutions of alternate cooligomers and block copolymers (i.e., $C_3H_7$-$(C_5H_{10}$-$C_3H_6)_4$-$C_5H_{11}$ and $C_3H_7$-$(C_3H_6)_9$-$(C_5H_{10})_9$-$C_5H_{11}$, respectively). Note that just as in earlier studies of RelRes that parameterize between the FG and CG potentials via a multipole approximation,[25-26] we have shown that the current implementation in LAMMPS retrieves the structural, thermal, static, dynamic, etc. behavior of the reference systems with negligible error. We thus expect that the RelRes algorithm can further capture free energies, heat transfers, etc. of various multifaceted processes (solvation, aggregation, etc.) in all mixtures governed by LJ interactions.

Interestingly, we have found that the proper switching distance for RelRes is generally driven by the mapping ratio. For example, for both propane and hexane (i.e., $C_3H_8$ and $C_3H_7$-$C_3H_7$, respectively), with a mapping ratio of 3, $r_s = 0.6$nm, yet for both neopentane and bineopentyl (i.e., $C_5H_{12}$ and $C_5H_{11}$-$C_5H_{11}$, respectively), with a mapping ratio of 5, $r_s = 0.7$nm. If one is interested in an even better representation of a reference system, a larger switching distance can be employed, while giving up some computational efficiency (e.g., moving from $r_s = 0.6$nm to $r_s = 0.7$nm costs about 20% of CPU time). Note that the current work restricts the mapping for groups whose ordinary sites are directly bonded to hybrid sites. Removing this limitation can favorably increase the mapping ratio but unfavorably increase the switching distance; exploring the computational efficiency of such alternatives becomes especially of interest.

Although the current implementation only deals with the LJ potential, it can be also used for systems with other potentials as well, just by invoking the `hybrid` pair style in LAMMPS. Moreover, there is the possibility to develop RelRes variations of the Coulomb potential (e.g., pair styles `coul/relres` or `lj/coul/relres` in the LAMMPS standard). This kind of an algorithm would be very useful for a group of atoms with a net charge (e.g., nitrate, ammonium, etc.); this would just require replacing $u^{FG}$ and $u^{CG}$ in Eq. (4) with the appropriate Coulomb expressions, in a similar manner as was done with Eqs. (11) and (12) for the LJ potential. However, for a group of atoms with no net charge (e.g., water), a more elaborate approach is necessary, since the zero order term of a multipole approximation (i.e., Eq. (4)) is not sufficient: Chaimovich et al. establishes the appropriate framework by deriving the first and second terms of the relevant Taylor series.[25-26] Implementing such functionalities would require employing the dipole in LAMMPS (e.g., pair styles `dipole/relres` or `lj/dipole/relres` in the LAMMPS standard). All these routes are very practical directions for future research, since they could all substantially speedup molecular simulations of polar mixtures.

In summary, we recommend our automated version of RelRes in LAMMPS for anyone studying molecular systems that involve LJ interactions. As compared with conventional simulations, this RelRes implementation can yield almost an order of magnitude enhancement in computational efficiency, whilst capturing the behavior of nonpolar mixtures with negligible error.

## Acknowledgements

We are grateful for all those who gave us moral and spiritual support in this project during the Coronavirus Pandemic.